\documentstyle[preprint,aps,epsfig]{revtex}

\tighten
\draft

\begin{document}
\preprint{\vbox{\hbox{PNU-NTG-01/2000} \hbox{RUB-TPII-04/00}}}

\title{Spin content of the $\Lambda$ hyperon}
\author{Hyun-Chul Kim$^a$, Micha{\l} Prasza{\l}owicz$^b$, and Klaus Goeke$^c$}
\address{~\\
$^a$ Department of Physics, Pusan National University,\\
Pusan 609-735, Republic of Korea \\
$^b$ Institute of Physics, Jagellonian University, \\
ul. Reymonta 4, 30-059 Krak{\'o}w, Poland \\
$^c$ Institute for Theoretical Physics II, Ruhr-University Bochum, \\
D-44780 Bochum, Germany}
\date{March, 2000}
\maketitle

\begin{abstract}
Using the known experimental data for the hyperon semileptonic decay
constants, we calculate integrated quark densities $\Delta q_\Lambda$ and $%
\Delta \Sigma_\Lambda$ for the hyperon $\Lambda$ with flavor SU(3) symmetry
breaking taken into account. Symmetry breaking is implemented with the help
of the chiral quark-soliton model in an approach, in which the dynamical
parameters are fixed by the experimental data for six hyperon semileptonic
decay constants. This parametrization allows us to reproduce the first
moment of the $g_1^{{\rm p}}(x)$ of the proton. For the $\Lambda$ we obtain: 
$\Delta u_{\Lambda}=\Delta d_{\Lambda} \approx 0$ and $\Delta s_{\Lambda}$
of the order of 1. Unfortunately large experimental errors of $\Xi^-$ decays
propagate in our analysis especially in the case of $\Delta\Sigma_{\Lambda}$
and $\Delta s_{\Lambda}$. Only if the errors for these decays are reduced,
the accurate theoretical predictions for $\Delta\Sigma_{\Lambda}$ and $%
\Delta s_{\Lambda}$ will be possible.
\end{abstract}


\section{Introduction}

The spin and flavor content of the nucleon has been an extensively studied
issue for well over a decade, since the EMC~\cite{EMC} announced that the
strange quark is strongly polarized opposite to the valence quarks,
indicating that the quarks carry only a small fraction of the nucleon spin.
In contrast to the Ellis-Jaffe sum rule \cite{EllisJaffe}, the polarization
of the strange quark turned out to be nonnegligible. A series of following
experiments confirmed the EMC result~\cite{SMC,E143,E154,HERMES1}. Hence, it
is also natural to investigate the structure of other baryons. In
particular, the $\Lambda $ hyperon is interesting, since $\Delta q_{\Lambda }
$ are related to the fragmentation functions which can be measured
experimentally \cite{BurkardtJaffe}. According to the naive quark model the
spin of $\Lambda $ comes solely from the strange quark, while the up and
down quarks form the spin singlet, so that they make no contribution.
However, an analysis based on the data of hyperon semileptonic decays and
lepton-nucleon deep-inelastic scattering~\cite
{BurkardtJaffe,Jaffe,BorosLiang} predicts $\Delta s_{\Lambda }\simeq 0.6$
and $\Delta u_{\Lambda }=\Delta d_{\Lambda }\simeq -0.2$. This analysis
assumes, however, the flavor SU(3) symmetry, {\em i.e.} it uses the
relations: $\Delta u=\Delta d=\frac{1}{3}(\Delta \Sigma -D)$ and $\Delta s=%
\frac{1}{3}(\Delta \Sigma +2D)$, with $\Delta \Sigma $ identical for all
octet baryons. It implies that the $\Lambda $ spin is not completely carried
by the strange quark. While the above analysis shows a discrepancy with the
naive quark model, one should note that the effect of SU(3) symmetry
breaking was not taken into account.

Recently, we have investigated hyperon semileptonic decays and the spin
content of the nucleon with SU(3) flavor symmetry breaking taken into
account~\cite{KimPraGo2}. The SU(3) symmetry breaking was implemented via
the Chiral Quark-Soliton Model ($\chi $QSM) in such a way that all dynamical
variables in the model were fixed by the experimental data for the
semileptonic decay constants. The results for the proton were: $\Delta u_{%
\text{p}}=0.72\pm 0.07$, $\Delta d_{\text{p}}=-0.54\pm 0.07$, $\Delta s_{%
\text{p}}=0.33\pm 0.51$, and $\Delta \Sigma _{\text{p}}=0.51\pm 0.41$. The
large errors in $\Delta s_{\text{p}}$ and $\Delta \Sigma _{\text{p}}$ are
due to the large experimental uncertainties of the $\Xi ^{-}$ decay
constants. The conclusion in that work was as follows: First, statements
concerning $\Delta s_{\text{p}}$ and $\Delta \Sigma _{\text{p}}$ based on
SU(3) flavor symmetry are premature. Second, accurate results can be
obtained only by reducing the experimental uncertainty for $\Xi $ decays. It
is of great interest to study the spin content of the $\Lambda $, using the
same framework as in Ref.~\cite{KimPraGo2}. The aim of this paper is thus to
find out what we can reliably conclude on the spin structure of the $\Lambda 
$, based on hyperon semileptonic decays.

Let us first briefly recall how the standard analysis is carried out. Three
diagonal axial-vector coupling constants define integrated polarized quark 
densities for a given baryon B: 
\begin{eqnarray}
g_{{\rm A}}^{(3)}(\text{B}) &=&\Delta u_{\text{B}}-\Delta d_{\text{B}}, 
\nonumber \\
\sqrt{3}g_{{\rm A}}^{(8)}(\text{B}) &=&\Delta u_{\text{B}}+\Delta d_{\text{B}
}-2\Delta s_{\text{B}},  \nonumber \\
g_{{\rm A}}^{(0)}(\text{B}) &=&\Delta u_{\text{B}}+\Delta d_{\text{B}
}+\Delta s_{\text{B}}.  \label{gA380}
\end{eqnarray}
Note that in our normalization $g_{{\rm A}}^{(0)}($B$)=$ $\Delta \Sigma _{%
\text{B}}$.

Assuming SU(3) symmetry, one can calculate $g_{\text{A}}^{(3,8)}($B$)$ in
terms of the reduced matrix elements $F$ and $D$. For the proton and $%
\Lambda $ one gets: 
\begin{eqnarray}
g_{{\rm A}}^{(3)}(\text{p})=F+D, &~~~ & \sqrt{3}g_{{\rm A}}^{(8)}(\text{p})
=3F-D,  \nonumber \\
g_{{\rm A}}^{(3)}(\Lambda )=0, &~~~ & \sqrt{3}g_{{\rm A}}^{(8)}(\Lambda)
=-2D.  \label{gA38FD}
\end{eqnarray}

The constants $F$ and $D$ can be in principle extracted from the
hyperon semileptonic decays. For example: 
\begin{equation}
A_{1}\;=\;\left( g_{1}/f_{1}\right) ^{({\rm n}\rightarrow {\rm p}%
)}=F+D\,,\;\;\;\;A_{4}\;=\;\left( g_{1}/f_{1}\right) ^{(\Sigma
^{-}\rightarrow {\rm n})}=F-D\,.  \label{A1A4exp}
\end{equation}
For convenience, we denote the ratios of axial-vector to vector decay
constants by $A_{i}$ (see Table I). Taking for these decays experimental
values (see Table I), one gets $F=0.46$ and $D=0.80$.

Since $g_{{\rm A}}^{(0)}($B$)$ does not correspond to the SU(3) current, it
cannot be expressed in terms of $F$ and $D$ without further assumptions.
Thus, in order to extract all $\Delta q_{\text{p}}$ separately, one needs
some additional information. Either another experimental input is needed, or
a model which predicts $g_{{\rm A}}^{(0)}($B$)$ in terms of the $F$ and $D$.

The first possibility can be realized by taking the experimental result for
the first moment of the spin structure function $g_{1}^{\text{p}}(x)$ of the
proton: 
\begin{equation}
I_{\text{p}}=\int\limits_{0}^{1}dx\,g_{1}^{\text{p}}(x)=\frac{1}{18}\left(
4\Delta u_{\text{p}}+\Delta d_{\text{p}}+\Delta s_{\text{p}}\right) \left( 1-%
\frac{\alpha _{\text{s}}}{\pi }+\ldots \right) .  \label{Ip0}
\end{equation}
Recent analysis~\cite{KarLip} implies $I_{\text{p}}=0.124\pm 0.011$ which
translates into: 
\begin{equation}
\Gamma _{\text{p}}\equiv 4\Delta u_{\text{p}}+\Delta d_{\text{p}}+\Delta s_{%
\text{p}}=2.56\pm 0.23\,.  \label{Gampval}
\end{equation}
if $\alpha _{{\rm s}}(Q^{2}=3~({\rm GeV}/c)^{2})=0.4$ is assumed. Taking for 
$F=0.46$ and for $D=0.80$ together with Eq.(\ref{Gampval}), one gets for the
nucleon: $\Delta u_{\text{p}}=0.79$, $\Delta d_{\text{p}}=-0.47$ and $\Delta
s_{\text{p}}=-0.13$, which implies $\Delta \Sigma _{\text{p}}=0.19$, a
fairly small number as compared with the naive expectation from the quark
model: $\Delta \Sigma _{\text{p}}=1$.

It is important to realize that $\Delta \Sigma _{\text{p}}$ is not directly
measured; it is extracted from the data through some theoretical model. In
the above example we have assumed the SU(3) symmetry and used two
arbitrarily chosen hyperon decays (\ref{A1A4exp}). One could, however, use
any two $A_{i}$'s out of 6 known hyperon decays to extract $F$ and $D.$ The
number of combinations which one can form to extract $F$ and $D$ is 14
(actually 15, but two conditions are linearly dependent). Taking these 14
combinations into account, one gets: 
\begin{equation}
F=0.40\div 0.55,\quad D=0.70\div 0.89\,.  \label{FDrange}
\end{equation}
These are the uncertainties of the {\em central values} due to the
theoretical error caused by using the exact SU(3) symmetry to describe the
hyperon semileptonic decays. These uncertainties are further increased by
the experimental errors of all individual decays.

Looking at Eq.(\ref{FDrange}), one might get an impression that a typical
error associated with using SU(3) symmetry in analyzing the hyperon decays
is of the order of 15 \% or so. While this is true for the hyperon decays,
the values of $\Delta q_{{\rm B}}$ and $\Delta \Sigma_{{\rm B}}$ of the
various baryons might be much more affected by the symmetry breaking. Indeed
for $F$ and $D$ corresponding to (\ref{FDrange}) and $\Gamma _{\text{p}}$ as
given by Eq.(\ref{Gampval}) $\Delta \Sigma _{\text{p}}=0.02\div 0.30$.

As will be shown in the following, the $\chi$QSM predicts
in the chiral limit~\cite{DiaPetPol}: 
\begin{equation}
g_{{\rm A}}^{(0)}(\text{B})=9F-5D  \label{ga0}
\end{equation}
for any baryon B. Here $g_{{\rm A}}^{(0)}$ is very sensitive to small
variations of $F$ and $D$, since it is a difference of the two, with
relatively large multiplicators. Indeed, for the 14 fits mentioned above the
central value for $g_{{\rm A}}^{(0)}$ of the nucleon varies between $-0.25$
to approximately 1 and a similar feature is expected for any baryon,
particularly for the $\Lambda $. Thus, despite the fact that hyperon
semileptonic decays are relatively well described by the model in the chiral
limit, the singlet axial-vector constant is basically undetermined. This is
a clear signal of the importance of the symmetry breaking for this quantity.

One could argue that this kind of behavior is just an artifact of the $\chi $
QSM. However, the scenario of a rotating soliton (which is by the way used
also in the Skyrme-type models) is very plausible and cannot be {\em a
priori } discarded on the basis of first principles. The $\chi $QSM is a
particular realization of this scenario and we use it as a tool to
investigate the sensitivity of the singlet axial-vector current to the
symmetry breaking effects in hyperon semileptonic decays. In fact,
conclusions similar to ours have been obtained in chiral perturbation theory
in Ref.\cite{SavWal}.

In Ref.~\cite{KimPraGo2} we have at length discussed the properties of the
model formula for $g_{{\rm A}}^{(0)}$ in two limiting cases , {\em i.e.}
large (Skyrme model limit) and small (quark model limit) soliton sizes. In
the Skyrme model the ratio $F/D=5/9$ and $\Delta \Sigma _{\text{p}}$
vanishes. In the quark model $F/D=2/3$ and $F+D=5/3$, and therefore $\Delta
\Sigma _{\text{p}}=1$. We also gave numerical arguments in support of our
approach: namely releasing the model assumptions concerning $g_{{\rm A}%
}^{(0)}$ and using $\Gamma _{\text{p}}$ as an additional input one arrives
at almost identical numerical results as using (\ref{ga0}).

It is virtually impossible to analyze the symmetry breaking in weak decays
without resorting to some specific model~\cite{KarLip}. In this paper,
following Ref.~\cite{KimPraGo2}, we will implement the symmetry breaking for
the hyperon decays using the $\chi $ QSM (see Ref.\cite{review} for review)
which satisfactorily describes the axial-vector properties of hyperons \cite
{BloPraGo}\nocite{BPG,Wakaspin}-- \cite{KimPoPraGo}.

The model provides a link between the matrix elements of the octet of the
axial-vector currents, responsible for hyperon decays, and the matrix
elements of the singlet axial-vector current. In the present work we will
study the relation between hyperon semileptonic decays and integrated polarized
quark distributions for the $\Lambda $ hyperon. We will use the $\chi $QSM
only to identify the algebraic structure of the symmetry breaking ($m_{\text{%
s}}$ corrections). The dynamical quantities, so called inertia parameters
which are in principle calculable within the model \cite{BloPraGo}, will be
treated as free parameters. By adjusting them to the experimentally known
semileptonic decays we allow not only for maximal phenomenological input but
also for minimal model dependence. In Ref.\cite{KimPraGo,strange} we have
already studied the magnetic moments of the octet and decuplet in this way.

Such a ''model-independent'' approach -- used for example by Adkins and
Nappi \cite{AdNap} in the context of the Skyrme model -- is of interest for
at least two reasons. First, it can be considered as a QCD-motivated tool
to analyze and classify (in terms of powers of $m_{{\rm s}}$ and $1/N_{{\rm c%
}}$) the symmetry-breaking terms for a given observable. For nontrivial
operators such as axial-vector form factors a general analysis, without
referring to some specific model, is virtually impossible. Second, this
''model-independent'' analysis provides an information for the model
builders as well. It tells us what are the best predictions the model can
ever produce. Indeed, model calculations in the framework of the $\chi $QSM
are not as unique as one might think: They depend on adopted regularizations,
cutoff parameter, or the constituent quark mass. Moreover, in the SU(3)
version of the $\chi $QSM a quantization ambiguity appears~\cite{paradox}.
Therefore, if the ``model-independent'' analysis would have failed to
describe the data, that would mean that the model did not correctly include
all necessary physics relevant for a given observable. On the other hand,
the success of such an analysis gives a strong hint for the model builders
that the model is correct and worth exploring.

As far as the symmetry breaking is concerned, our results are identical to
the ones obtained in Refs.\cite{Man} within QCD in the large $N_{{\rm c}}$
limit. Indeed, the $\chi $QSM is a specific realization of the large $N_{%
{\rm c}}$ limit. The truly new ingredient of our analysis is the model
formula for the singlet axial-vector constant $g_{{\rm A}}^{(0)}$, {\em i.e.}
Eq.(\ref{ga0}), which we use to calculate quantities relevant for polarized
high energy experiments.

\section{Hyperon decays in the Chrial Quark Soliton Model}

The discussion in this section follows closely Ref.~\cite{KimPraGo2}. The
transition matrix elements of the hadronic axial-vector current $\langle
B_2|A_\mu ^X|B_1\rangle $ can be expressed in terms of three independent
form factors: 
\begin{equation}
\langle B_2|A_\mu ^X|B_1\rangle \;=\;\bar{u}_{B_2}(p_2)\left[ \left\{
g_1^{B_1\rightarrow B_2}(q^2)\gamma _\mu -\frac{ig_2^{B_1\rightarrow
B_2}(q^2)}{M_1}\sigma _{\mu \nu }q^\nu +\frac{g_3^{B_1\rightarrow B_2}(q^2)}{
M_1}q_\mu \right\} \gamma _5\right] u_{B_1}(p_1),
\end{equation}
where the axial-vector current is defined as 
\begin{equation}
A_\mu ^X\;=\;\bar{\psi}(x)\gamma _\mu \gamma _5\lambda _X\psi (x)
\label{Eq:current}
\end{equation}
with $X=\frac 12(1\pm i2)$ for strangeness conserving $\Delta S=0$ currents
and $X=\frac 12(4\pm i5)$ for $|\Delta S|=1$. Similar expressions hold for
the hadronic vector current, where the $g_i$ are replaced by $f_i$ ($i=1,2,3$
) and $\gamma _5$ by ${\bf 1}$.

Hadronic matrix elements such as $\langle B_{2}|A_{\mu }^{X}|B_{1}\rangle $
can be easily evaluated within the $\chi $QSM~\cite{review}. 
Taking into account the $1/N_{c}$ rotational and $%
m_{{\rm s}}$ corrections, we can write the resulting axial-vector constants $%
g_{1}^{B_{1}\rightarrow B_{2}}(0)$ in the following form\footnote{%
In the following we will assume that the baryons involved have $S_{3}=+\frac{%
1}{2}$.}: 
\begin{eqnarray}
g_{1}^{(B_{1}\rightarrow B_{2})} &=&a_{1}\langle
B_{2}|D_{X3}^{(8)}|B_{1}\rangle \;+\;a_{2}d_{pq3}\langle B_{2}|D_{Xp}^{(8)}\,%
\hat{S}_{q}|B_{1}\rangle \;+\;\frac{a_{3}}{\sqrt{3}}\langle
B_{2}|D_{X8}^{(8)}\,\hat{S}_{3}|B_{1}\rangle   \nonumber \\
&+&m_{s}\left[ \frac{a_{4}}{\sqrt{3}}d_{pq3}\langle
B_{2}|D_{Xp}^{(8)}\,D_{8q}^{(8)}|B_{1}\rangle +a_{5}\langle B_{2}|\left(
D_{X3}^{(8)}\,D_{88}^{(8)}+D_{X8}^{(8)}\,D_{83}^{(8)}\right) |B_{1}\rangle
\right.   \nonumber \\
&+&\left. a_{6}\langle B_{2}|\left(
D_{X3}^{(8)}\,D_{88}^{(8)}-D_{X8}^{(8)}\,D_{83}^{(8)}\right) |B_{1}\rangle 
\right] .  \label{Eq:g1}
\end{eqnarray}
$\hat{S}_{q}$ ($\hat{S}_{3}$) stand for the $q$-th (third) component of the
spin operator of the baryons. The $D_{ab}^{({\cal R})}$ denote the SU(3)
Wigner matrices in representation ${\cal R}$. The $a_{i}$ denote parameters
depending on the specific dynamics of the chiral soliton model (see for 
example Refs.~\cite{review,SW} and references therein). 
Their explicit form in terms of a Goldstone mean field can be found in 
Ref.~\cite {BloPraGo}. As mentioned already, in the present approach we will 
not calculate this mean field but treat $a_{i}$ as free parameters to be
adjusted to experimentally known semileptonic hyperon decays.

Because of the SU(3) symmetry breaking due to the strange quark mass $m_{%
{\rm s}}$, the collective baryon Hamiltonian is no more SU(3)-symmetric. The
octet states are mixed with the higher representations such as antidecuplet $%
\overline{{\bf 10}}$ and eikosiheptaplet ${\bf 27}$~\cite{KimPraGo}. In the
linear order in $m_{{\rm s}}$ the wave function of a state $B=(Y,I,I_3)$ of
spin $S_3$ is given as: 
\begin{equation}
\psi _{B,S_3}=(-)^{\frac 12-S_3}\left( \sqrt{8}\,D_{B\,S}^{(8)}+c_B^{(%
\overline{10})}\sqrt{10}\,D_{B\,S}^{(\overline{10})}+c_B^{(27)}\sqrt{27}%
\,D_{B\,S}^{(27)}\right) ,
\end{equation}
where $S=(-1,\frac 12,S_3)$. Mixing parameters $c_B^{({\cal R})}$ can be
found for example in Ref.~\cite{BloPraGo}. They are given as products of a
numerical constant $N_B^{({\cal R})}$ depending on the quantum numbers of
the baryonic state $B$ and a dynamical parameter $c_{{\cal R}}$ depending
linearly on $m_{{\rm s}}$ (which we assume to be 180~MeV) and the model
parameter $I_2$, which is responsible for the splitting between the octet
and higher exotic multiplets~\cite{DiaPetPol,timeord}.

Analogously to Eq.(\ref{Eq:g1}), one obtains in the $\chi $QSM diagonal
axial-vector coupling constants. In that case $X$ can take two values: $X=3$
and $X=8$. For $X=0$ (singlet axial-vector current) we have the following
expression \cite{BloPraGo,BPG}: 
\begin{equation}
\frac 12\,g_A^{(0)}(\text{B})=\frac 12a_3+\sqrt{3}\,m_{\text{s}%
}\,(a_5-a_6)\;\langle B|D_{83}^{(8)}|B\rangle .  \label{Eq:singlet}
\end{equation}

This equation is remarkable, since it provides a link between an octet and
singlet axial-vector current. It is the most important model input in our
analysis. Pure QCD-arguments based the large $N_c$ expansion~\cite{Man} do
not provide such a link. Moreover, due to the structure of the matrix
element $\;\langle B|D_{83}^{(8)}|B\rangle $, the $g_A^{(0)}($B$)$ are
identical inside the isospin multiplets. We predict much stronger symmetry
breaking for the $\Lambda$ than for the proton, since 
\begin{equation}
\sqrt{3}\langle p|D_{83}^{(8)}|p\rangle =-\frac 1{10},\quad \sqrt{3}\langle
\Lambda |D_{83}^{(8)}|\Lambda \rangle =\frac 3{10}\,,  \label{D83}
\end{equation}
for spin $S_3=+1/2$.

Instead of calculating 7 dynamical parameters $a_i(i=1,\cdots ,6)$ and $I_2$
(which enters into $c_{\overline{10}}$ and $c_{27}$) within the $\chi $QSM,
we shall fit them from the hyperon semileptonic decays data. It is
convenient to introduce the following set of 7 new parameters: 
\[
r=\frac 1{30}\left( a_1-\frac 12a_2\right) ,\;\;\;\;\;\;s=\frac 1{60}%
a_3,\;\;\;x=\frac 1{540}m_{{\rm s}}\,a_4,\;\;\;y=\frac 1{90}m_{{\rm s}%
}\,a_5,\;\;\;z=\frac 1{30}m_{{\rm s}}\,a_6, 
\]
\begin{equation}
p=\frac 16m_{{\rm s}}\,c_{\overline{10}}\left( a_1+a_2+\frac 12a_3\right)
,\;\;\;q=-\frac 1{90}m_{{\rm s}}\,c_{27}\left( a_1+2a_2-\frac 32a_3\right) .
\label{Eq:newp}
\end{equation}

Employing this new set of parameters, we can express all possible
semileptonic decays of the octet baryons: 
\begin{eqnarray}
A_1\;=\;\left( {g_1}/{f_1}\right) ^{({\rm n}\rightarrow {\rm p})}
&=&-14r+2s-44x-20y-4z-4p+8q,  \nonumber \\
A_2\;=\;\left( {g_1}/{f_1}\right) ^{(\Sigma ^{+}\rightarrow \Lambda )}
&=&-9r-3s-42x-6y-3p+15q,  \nonumber \\
A_3\;=\;\left( {g_1}/{f_1}\right) ^{(\Lambda \rightarrow {\rm p})}
&=&-8r+4s+24x-2z+2p-6q,  \nonumber \\
A_4\;=\;\left( {g_1}/{f_1}\right) ^{(\Sigma ^{-}\rightarrow {\rm n})}
&=&4r+8s-4x-4y+2z+4q,  \nonumber \\
A_5\;=\;\left( {g_1}/{f_1}\right) ^{(\Xi ^{-}\rightarrow \Lambda )}
&=&-2r+6s-6x+6y-2z+6q,  \nonumber \\
A_6\;=\;\left( {g_1}/{f_1}\right) ^{(\Xi ^{-}\rightarrow \Sigma ^0)}
&=&-14r+2s+22x+10y+2z+2p-4q.  \label{Eq:semilep}
\end{eqnarray}
The U(1) and SU(3) axial-vector constants $g_A^{(0,3,8)}$ can be also
expressed in terms of the new set of parameters (\ref{Eq:newp}). In the case
of the proton and the $\Lambda$ we have the singlet axial-vector constants: 
\begin{equation}
g_A^{(0)}({\rm p})=60s-18y+6z,\hspace{0.5cm}g_A^{(0)}(\Lambda )=60s+54y-18z,
\label{Eq:singlet1}
\end{equation}
for the triplet ones, we write\footnote{%
Triplet $g_{A}^{(3)}$'s are proportional to $I_3$, formulae in Eq.(\ref
{Eq:triplet}) correspond to the highest isospin state.}: 
\begin{equation}
g_A^{(3)}({\rm p})=-14r+2s-44x-20y-4z-4p+8q,\hspace{0.5cm}g_A^{(3)}({\Lambda 
})=0,  \label{Eq:triplet}
\end{equation}
and for the octet one, we obtain: 
\begin{equation}
g_A^{(8)}({\rm p})=\sqrt{3}(-2r+6s+12x+4p+24q),\hspace{0.5cm}g_A^{(3)}({%
\Lambda })=\sqrt{3}(6r+2s-36x+36q).  \label{Eq:octet}
\end{equation}

Let us finally note that there is certain redundancy in Eq.(\ref{Eq:semilep}-%
\ref{Eq:octet}), namely by redefinition of $q$ and $x$ we can get rid of the
variable $p$: 
\begin{equation}
x^{\prime }=x-\frac{1}{9}p,\;\;\;\;q^{\prime }=q-\frac{1}{9}p.
\end{equation}

\section{Spin content of $\Lambda $ hyperon}

As shown in the last section there are 6 free parameters which have to be
fitted from the data. There are 2 {\em chiral} parameters: $r$ and $s$,
related closely to $F$ and $D$: 
\begin{equation}
F=5(s-r),\quad D=-3(s+3r).  \label{FDrs}
\end{equation}
and 4 proportional to $m_{{\rm s}}$: $x^{\prime }$, $y$, $z$, and $q^{\prime
}$. Since there are six known hyperon semileptonic decays, we can express 
all model parameters as linear combinations of these decay constants, and 
subsequently all quantities of interest can be expressed in terms of the input
amplitudes. In the following we will use the experimental values of Refs.~ 
\cite{PDG96,BGHORS}, which are presented in Table I.

Before doing this, let us, however, observe that there exist two linear
combinations $A_{i}$'s which within the model are free of the $m_{{\rm s}}$
corrections: 
\begin{eqnarray}
-42r+6s &=&\,A_{1}+2A_{6},  \nonumber \\
90r+90s &=&3A_{1}-8A_{2}-6A_{3}+6A_{4}+6A_{5}\,.  \label{rs0}
\end{eqnarray}
Solving Eq.(\ref{rs0}) for $r$ and $s$, we obtain the {\em chiral-limit}
expressions for hyperon semileptonic decays and integrated quark densities 
(i.e. with $x^{\prime }=y=z=q^{\prime }=0$).  The corresponding $F$ and $D$ 
take the following form: 
\begin{eqnarray}
F &=&\frac{1}{12}(4A_{1}-4A_{2}-3A_{3}+3A_{4}+3A_{5}+5A_{6}),  \nonumber \\
D &=&\frac{1}{12}(4A_{2}+3A_{3}-3A_{4}-3A_{5}+3A_{6}).  \label{FDAi}
\end{eqnarray}
Numerically: 
\begin{equation}
F=0.50\pm 0.07,\quad D=0.77\pm 0.04.  \label{FDchiral}
\end{equation}
With these values for $F$ and $D$ together with Eq.(\ref{Gampval}) one gets: 
$\Delta u_{\text{p}}=0.81$, $\Delta d_{\text{p}}=-0.47$ and $\Delta s_{\text{%
p}}=-0.20$, which implies $\Delta \Sigma _{\text{p}}=0.15$. The advantage of
using Eq.(\ref{FDAi}) consists in the fact that $F$ and $D$ do not need to
be refitted when $m_{\text{s}}$ corrections are added.

Another important point is, that Eq.(\ref{FDAi}) is more general than the
model considered here. In fact they follow from the large $N_{\text{c}}$
QCD, as discussed in Ref.~\cite{Man}. The errors come from the experimental
errors of the decay amplitudes and are dominated by the errors of the $\Xi
^{-}$ decays. It is of utmost importance to reduce the errors of these
decays in order to get better accuracy for $F$ and $D$.

In the case of the $\Lambda$ Eq.(\ref{gA38FD}) implies that $\Delta u_{\Lambda
}=\Delta d_{\Lambda }$ and one has in the chiral limit: 
\begin{eqnarray}
\Delta u_{\Lambda }^{(0)} &=&3F-2D=A_{1}-\frac{5}{3}A_{2}-\frac{5}{4}A_{3}+%
\frac{5}{4}A_{4}+\frac{5}{4}A_{5}+\frac{3}{4}A_{6\,,}  \nonumber \\
\Delta s_{\Lambda }^{(0)} &=&3F-D=A_{1}-\frac{4}{3}%
A_{2}-A_{3}+A_{4}+A_{5}+A_{6\,.}  \label{uds0L}
\end{eqnarray}
Numerical values $\Delta u_{\Lambda }^{(0)}=\Delta d_{\Lambda
}^{(0)}=-0.03\pm 0.14$ and $\Delta s_{\Lambda }^{(0)}=0.74\pm 0.17$ (see
Table II) are closer to the naive quark model expectations: 
$\Delta u_\Lambda =\Delta d_{\Lambda }=0$ and $\Delta s_{\Lambda }=1$, 
than to the numbers quoted in
Ref.\cite{BurkardtJaffe,Jaffe}: $\Delta u_{\Lambda }=\Delta d_{\Lambda
}=-0.23\pm 0.06$ and $\Delta s_{\Lambda }=0.58\pm 0.07$. This is reflected
in the fact that 
\begin{equation}
\Delta \Sigma ^{(0)}=9F-5D=3A_{1}-\frac{14}{3}A_{2}-\frac{7}{2}A_{3}+\frac{7%
}{2}A_{4}+\frac{7}{2}A_{5}+\frac{5}{2}A_{6}\,.  \label{GL0}
\end{equation}
(which is identical to all hadrons) reads: $\Delta \Sigma ^{(0)}=0.68\pm 0.44
$ and is much larger than the value required by using $\Gamma _{\text{p}}$
as an additional input. Indeed, as explained in Ref.~\cite{KimPraGo2}, in
the chiral limit one is not able to reproduce the value of $\Gamma _{\text{p}%
}$ (see Table II).

The two least known amplitudes $A_5$ and $A_6$ are almost entirely
responsible for the errors of $\Delta q_\Lambda $. However, since the
coefficients which enter into Eqs.(\ref{uds0L},\ref{GL0}) are not too large,
the absolute errors are relatively small.

The full expressions are obtained by solving the remaining 4 equations for $%
m_{{\rm s}}$ dependent parameters $x^{\prime }$, $y$, $z$ and $q^{\prime }$.
Also in this case we are able to link integrated quark densities $\Delta q$
to the hyperon decays: 
\begin{eqnarray}
\Delta u_\Lambda &=& \Delta d_\Lambda \;=\; -{\frac{\,{A_2}}3}-{\frac{\,{A_3}%
}4}+{\frac{\,{A_4}}4}+{\frac{13{A_5}}4}-{\frac{{A_6}}4}\,,  \nonumber \\
\Delta s_\Lambda &=&{\frac{15\,{A_1}}4}-{\frac{13\,{A_2}}2}-{\frac{87\,{A_3}%
}{16}-\frac{21\,{A_4}}{16}}+{\frac{45\,{A_5}}{16}}+{\frac{51\,{A_6}}{16}} 
\nonumber \\
\Delta \Sigma _\Lambda &=&\frac{15\,{A_1}}4-{\frac{46{A_2}}6}-{\frac{95\,{A_3%
}}{16}}-{\frac{13\,{A_4}}{16}}+{\frac{149\,{A_5}}{16}}+{\frac{46\,{A_6}}{16}}%
\,.  \label{delLam}
\end{eqnarray}

To guide the eye it is convenient to restore the linear $m_{\text{s}}$
dependence for the quark densities in the following way: 
\[
\Delta q=\Delta q^{(0)}+\frac{m_{\text{s}}}{180\,\text{MeV}}\left( \Delta
q-\Delta q^{(0)}\right) ,
\]
and similarly for $\Delta \Sigma $. This dependence is explicitly shown in
Fig.1, where we plot the central values and ``experimental'' error bars
(shaded areas) of $\Delta q_{\Lambda }$'s. 

\vspace{0.5cm} \centerline{\epsfysize=2.7in\epsffile{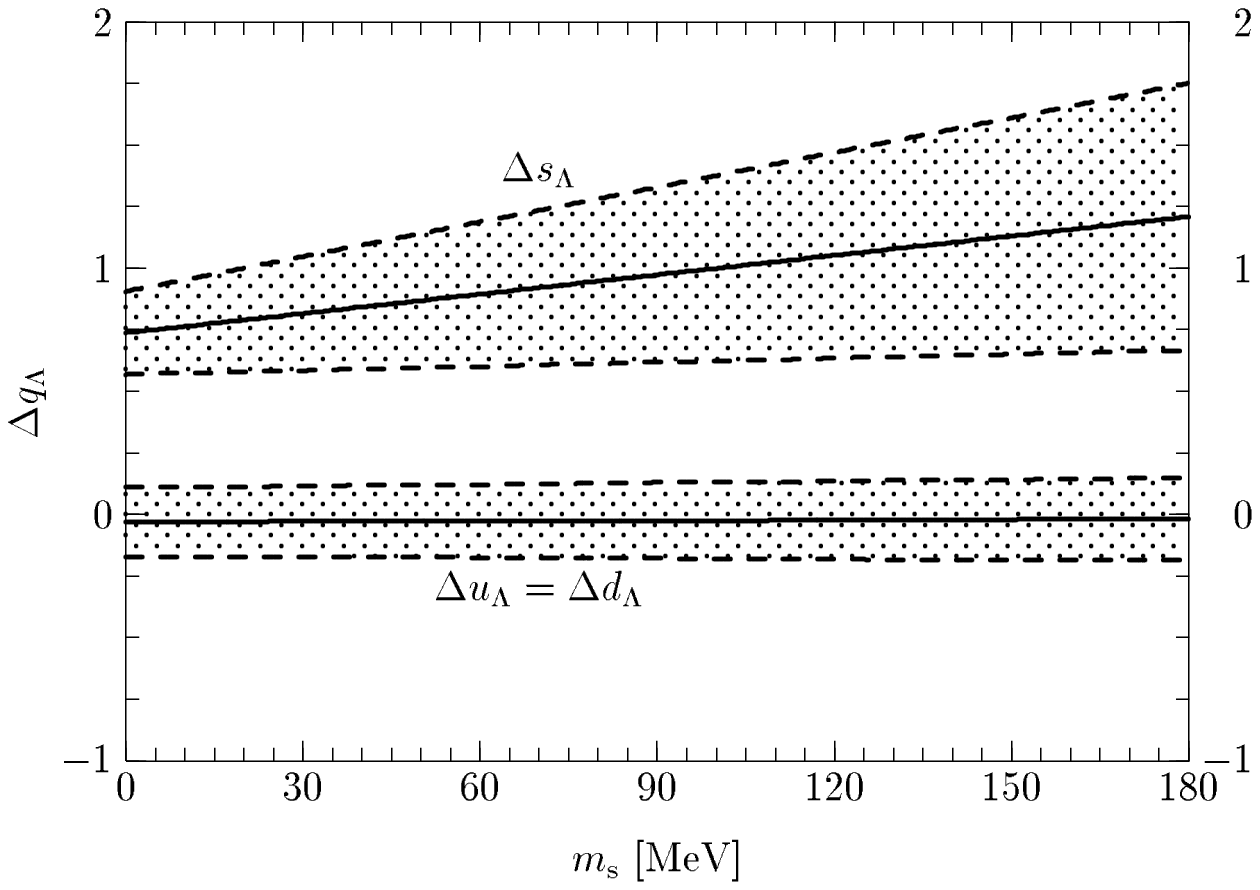}} \vskip%
4pt 
\centerline{\ Fig.1.~$\Delta q_{\Lambda}$ as functions of $m_{{\rm s}}$. 
} \vspace{0.5cm} 

\vspace{0.5cm}
\centerline{\epsfysize=2.7in\epsffile{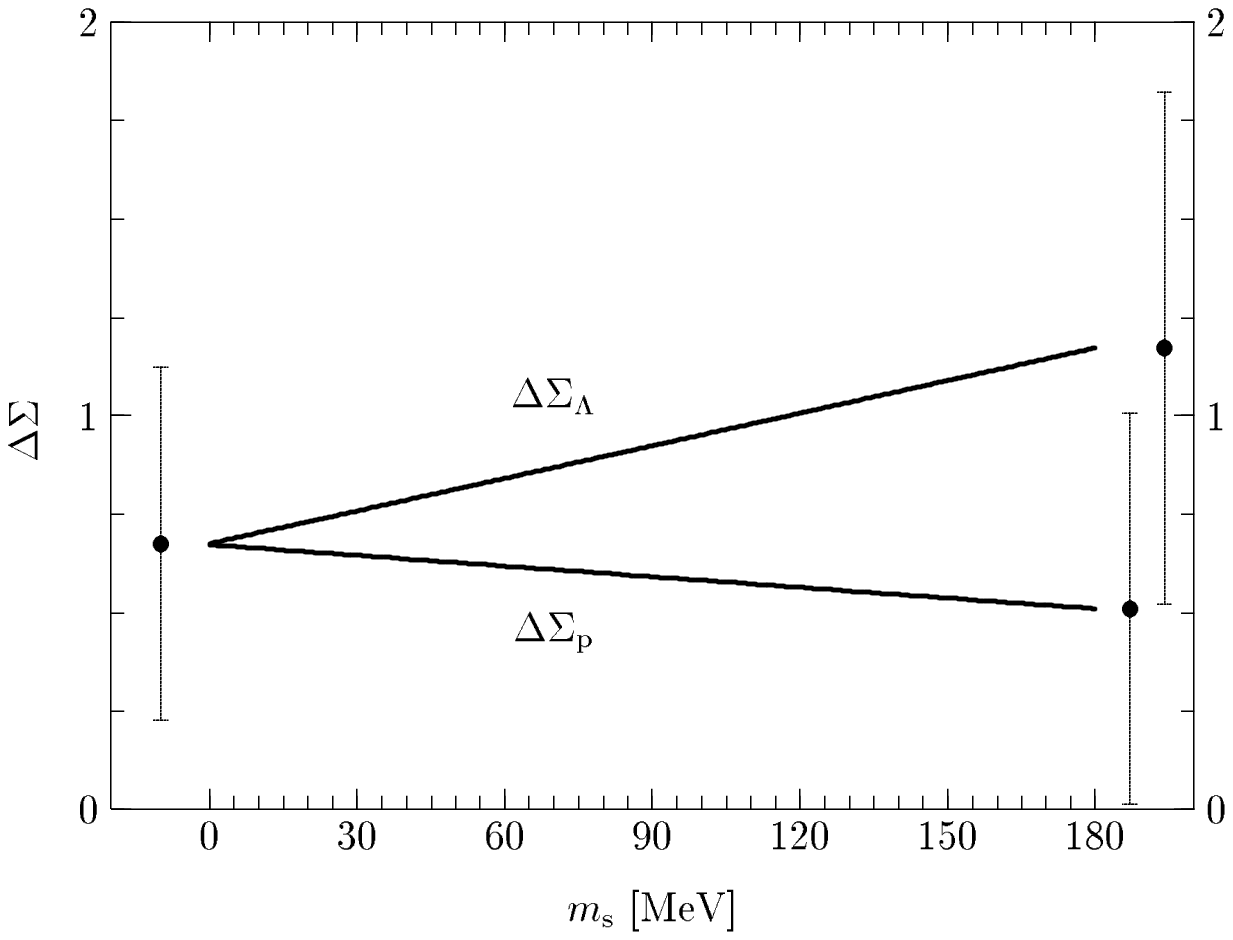}} \vskip4pt 
\centerline{\
Fig.2.~$\Delta \Sigma_{{\rm p}}$ and $\Delta \Sigma_{\Lambda}$ as functions
of $m_{{\rm s}}$. } \vspace{0.5cm} 

In Fig.2 we plot $m_{\text{s}}$
dependence of $\Delta \Sigma $ both for the proton and for the $\Lambda $.
In order to make the plot readable we have denoted theoretical errors as
error bars around the black dots which correspond to the chiral limit and
full theoretical prediction. The splitting between the proton and the 
$\Lambda $ is caused by the term proportional to $a_{5}-a_{6}$ in 
Eq.(\ref{Eq:singlet}).  Numerical values can be found in Table II. 
We see that for $m_{\text{s}}=180$
MeV apart from fitting all hyperon semileptonic decays (which is our input) 
we reproduce $\Gamma _{\text{p}}$ with relatively small error. The errors of 
$\Delta\Sigma $ and $\Delta s$ are much bigger. The central values, however, 
differ from the ``standard'' ones. Interestingly $\Delta s_{\text{p}}$ in 
proton is rather large and positive, however, the error bars are so large 
that the quark model value $\Delta s_{\text{p}}=0$ is not excluded. In the 
$\Lambda$ the $\Delta s_{\Lambda }$ is larger than 1, but again the errors 
are large.  The errors for $\Delta u$ and $\Delta d$ both in the proton and 
in the $\Lambda$ are much smaller.  For the $\Lambda$ we get that the 
non-strange quarks almost do not carry spin in surprising accordance with the 
expectations of the naive quark model.

As already discussed, the errors on for $\Delta q$'s and $\Delta \Sigma $
come almost entirely from the large errors of the $\Xi ^{-}$ decays ($A_5$
and $A_6$). Instead of using these two hyperon semileptonic decays $A_5$ and 
$A_6$ as input, we can use the experimental value for $\Gamma _{\text{p}}$ as 
given by Eq.(\ref{Gampval}) and $\Delta \Sigma _{\text{N}}$, which we vary in 
the range from 0 to 1. In Fig.3 we plot our predictions for $A_5$ and $A_6$
(solid lines), together with the experimental error bands for these two
decays. It is clearly seen from Fig.3 that the allowed region for $\Delta
\Sigma _{\text{N}}$, in which the theoretical prediction falls within the
experimental error bars amounts to $\Delta \Sigma _{\text{N}}=0.20\div 0.45$.

\vspace{0.5cm} \centerline{\epsfysize=2.7in\epsffile{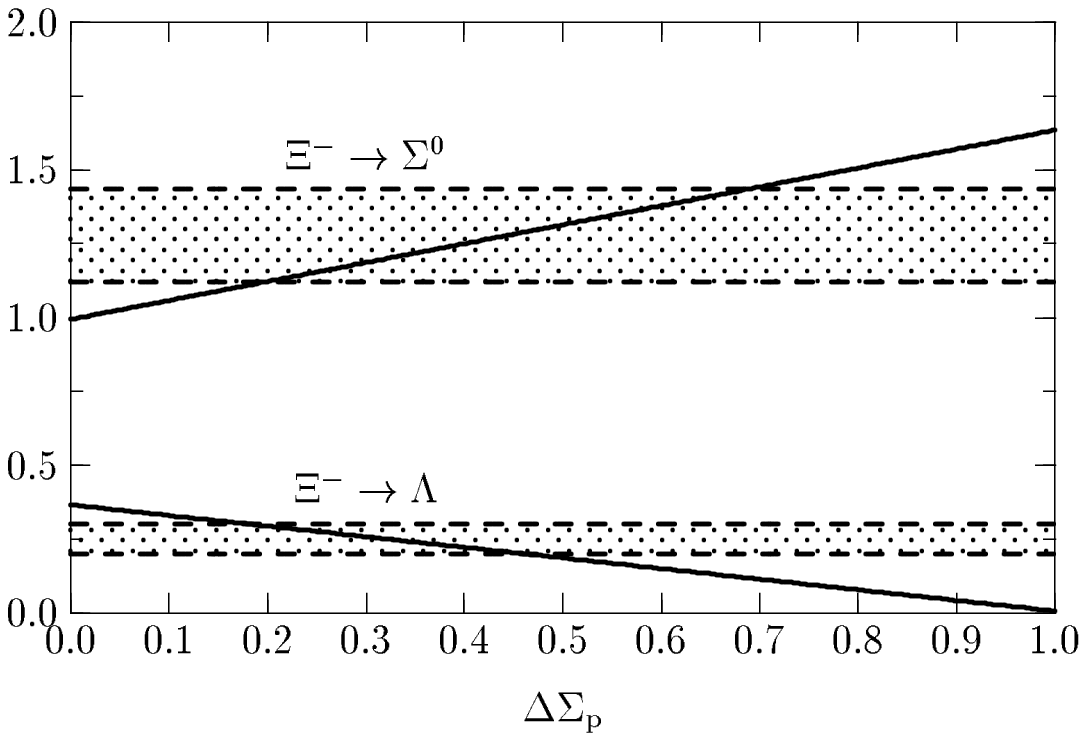}} \vskip%
4pt 
\centerline{\ Fig.3.~$A_5$ (lower line) and $A_6$ (upper line) as
functions of $\Delta \Sigma_{{\rm p}}$. } \vspace{0.5cm} 

\vspace{0.5cm} \centerline{\epsfysize=2.7in\epsffile{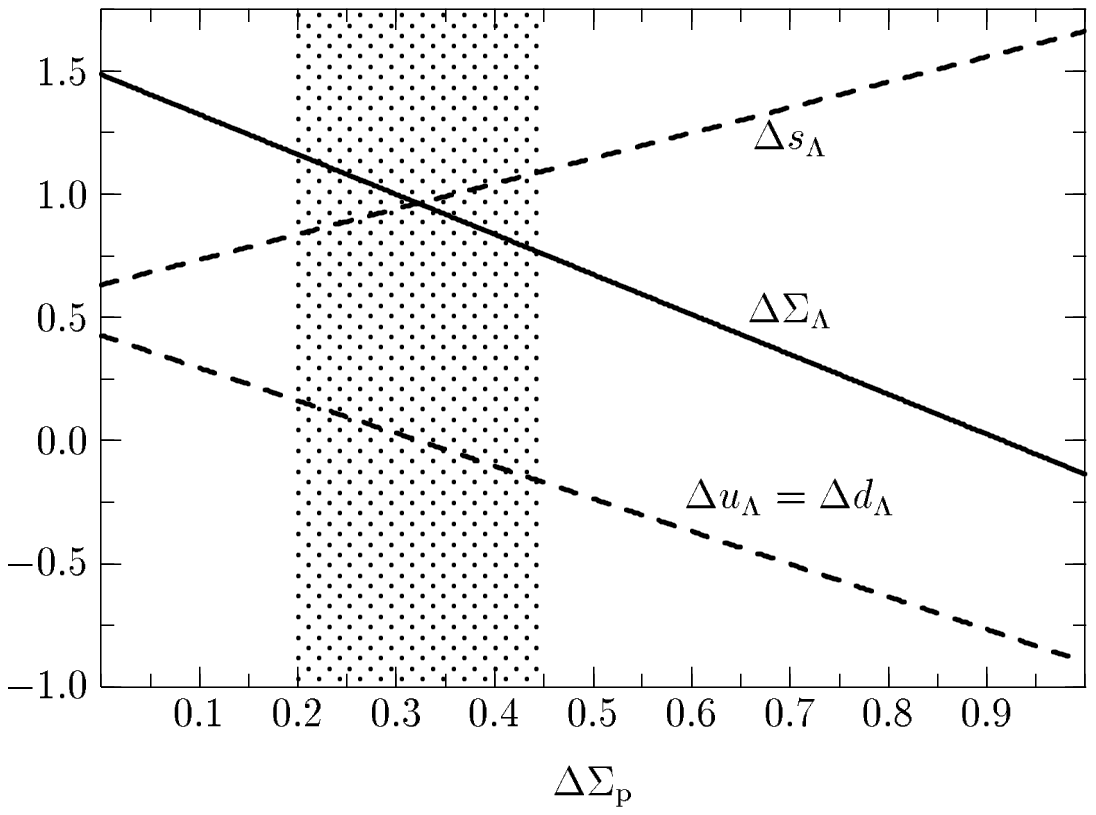}} \vskip%
4pt 
\centerline{\ Fig.4.~$\Delta q_{\Lambda}$'s and $\Delta\Sigma_{\Lambda}$
as functions of $\Delta \Sigma_{{\rm p}}$. } \vspace{0.5cm} 

In Fig.4 we plot the dependence of $\Delta q_{\Lambda }$'s and $\Delta
\Sigma _{\Lambda }$ upon $\Delta \Sigma _{\text{p}}$ (with $\Gamma _{\text{p}%
}$ fixed by Eq.(\ref{Gampval})). We see rather strong correlation of these
quantities with $\Delta \Sigma _{\text{p}}$. Within the allowed region $%
0.20<\Delta \Sigma _{\text{p}}<0.45$ the strange quark density $\Delta
s_{\Lambda }$ varies from 0.84 to 1.10. Interestingly, in the central region
around $\Delta \Sigma _{\text{p}}\approx 0.35$ the strange quark density in $%
\Lambda $ is close to 1 in accordance with an intuitive assumptions of the
naive quark model. Nonstrange quarks contribute to the spin of the Lambda at
the level of $-0.04$, and $\Delta \Sigma _{\Lambda }=0.92$.

\section{Summary}

In this paper we studied the influence of the SU(3) symmetry breaking in 
hyperon semileptonic decays on the determination of the integrated polarized
quark densities $\Delta q_\Lambda$ in the $\Lambda $. Using the Chiral
Quark-Soliton Model we have obtained a satisfactory parametrization of all
available experimental data on semileptonic decays. In this respect our
analysis is identical to QCD analysis in the large $N_{{\rm c}}$ of Ref.\cite
{Man}.

The new ingredient of our analysis consists in using the model formula for
the singlet axial-vector current in order to make contact with the high
energy polarization experiments.

The model contains 6 free parameters which can be fixed by 6 known hyperon
decays. Unfortunately $g_1/f_1$ for the two known decays of $\Xi ^{-}$ have
large experimental errors, which influence our predictions for $\Delta
q_\Lambda$. Our strategy was very simple: using model parametrization we
expressed $\Delta q_\Lambda$'s and $\Delta \Sigma_\Lambda$ in terms of the
six known hyperon decays. Errors were added in quadrature.

There are two points which have to be stressed here. Our fit respects chiral
symmetry in a sense that the leading order parameters $r$ and $s$ (or
equivalently $F$ and $D$) are fitted to the linear combinations of the
hyperon decays which are free from $m_{{\rm s}}$ corrections. As discussed
in Ref.~\cite{KimPraGo2} it is impossible to use the SU(3) symmetric
parametrization as given by Eq.(\ref{rs0}) and reproduce $\Gamma _{\text{p}}$
(as far as the central values are concerned) . With $m_{{\rm s}}$
corrections turned on one hits the experimental value for $\Gamma _{\text{p}%
} $ (see Table II), however, the value of $\Delta \Sigma _{\text{p}}$ is
practically undetermined, due the the experimental error of $\Xi ^{-}$
decays.

The nature of the $m_{\text{s}}$ is such that the central value of $\Delta
\Sigma _{\text{p}}$ is relatively large, whereas $\Delta s_{\text{p}}$ is
positive, however, still compatible with 0 within large errors. So one can
accommodate all existing data with $\Delta q_{\text{p}}$ much closer to the
expectations of the naive quark model than in the standard, SU(3) symmetric
approach. This trend is even stronger in the case of the $\Lambda $, where 
$\Delta u_{\Lambda }=\Delta d_{\Lambda }\approx 0$ and $\Delta s_{\Lambda
}\approx 1$. SU(3) symmetry breaking effects cause $\Delta \Sigma _{\Lambda
}>\Delta \Sigma _{\text{p}}$, so that the $\Lambda $ is in a sense more
nonrelativistic than the nucleon.

Our analysis shows clearly that if one wants to link hyperon semileptonic
decays with high-energy polarized experiments, one cannot neglect SU(3)
symmetry breaking for the former. In this respect our conclusion agrees with
Refs.\cite{EhrSch,SavWal}. Similarly to Ref.\cite{EhrSch} we see that $%
\Delta s_{\text{p}}=0$ is not ruled out by present experiments. Therefore,
the results for $\Delta s_\Lambda$ and $\Delta \Sigma_\Lambda $ 
based on the exact SU(3) symmetry are in our opinion premature.

\section*{Acknowledgments}

This work has partly been supported by the BMBF, the DFG and the
COSY--Project (J\"{u}lich). We are grateful to M.V. Polyakov for fruitful
discussions. H.-Ch.K. and M.P. thank P.V. Pobylitsa for critical comments.
H.-Ch.K. has been supported by the Korean Physical Society (1999). M.P. has
been supported by Polish grant {PB~2~P03B~019~17}.

\newpage


\newpage

\begin{table}[tbp]
\caption{The parameters $r, \ldots, q^\prime$ fixed to the experimental data
of the semileptonic decays \protect\cite{PDG96,BGHORS} $A_1$ -- $A_6$. The
entries for $A_1$ -- $A_6$ for the full fit (last column) correspond to the
experimental data.}
\begin{tabular}{cccc}
&  & chiral limit & with $m_{{\rm s}}$ \\ \hline
& $r$ & $-0.0892 $ & $-0.0892 $ \\ 
& $s$ & $0.0113 $ & $0.0113 $ \\ 
& $x^{\prime}$ & $0~~~~ $ & $-0.0055 $ \\ 
& $y$ & $0~~~~ $ & $0.0080 $ \\ 
& $z$ & $0~~~~ $ & $-0.0038 $ \\ 
& $q^{\prime}$ & $0~~~~ $ & $-0.0140 $ \\ \hline
$A_1$ & $\left({g_1}/{f_1}\right)^{n\rightarrow p}$ & $1.271\pm 0.11$ & $%
1.2573\pm 0.0028$(Input) \\ 
$A_2$ & $\left({g_1}/{f_1}\right)^{\Sigma^+\rightarrow \Lambda}$ & $0.769\pm
0.04$ & $0.742 \pm 0.018~~ $(Input) \\ 
$A_3$ & $\left({g_1}/{f_1}\right)^{\Lambda\rightarrow p}$ & $0.758\pm 0.08$
& $0.718 \pm 0.015~~ $(Input) \\ 
$A_4$ & $\left({g_1}/{f_1}\right)^{\Sigma^-\rightarrow n}$ & $-0.267\pm0.04$
& $-0.340 \pm 0.017~~ $(Input) \\ 
$A_5$ & $\left({g_1}/{f_1}\right)^{\Xi^-\rightarrow \Lambda}$ & $0.246\pm
0.07$ & $0.25 \pm 0.05~~~ $(Input) \\ 
$A_6$ & $\left({g_1}/{f_1}\right)^{\Xi^-\rightarrow \Sigma^0}$ & $1.271\pm
0.11$ & $1.278 \pm 0.158~~ $(Input)
\end{tabular}
\end{table}

\begin{table}[tbp]
\caption{Integrated quark densities $\Delta q$ and $\Delta\Sigma$ for the
nucleon (Ref.[??]) and for $\Lambda$. }
\begin{tabular}{ccrr}
&  & chiral limit & with $m_{{\rm s}}$ \\ \hline
& $\Delta u_{{\rm p}}$ & $0.98\pm 0.23$ & $0.72 \pm 0.07$ \\ 
& $\Delta d_{{\rm p}}$ & $-0.29\pm 0.13$ & $-0.54\pm 0.07$ \\ 
& $\Delta s_{{\rm p}}$ & $-0.02\pm 0.09$ & $0.33\pm 0.51$ \\ 
& $\Delta\Sigma_{{\rm p}}$ & $0.68 \pm 0.44$ & $0.51 \pm 0.41 $ \\ 
& $\Gamma_{{\rm p}}$ & $3.63 \pm 1.12$ & $2.67 \pm 0.33 $ \\ \hline
& $\Delta u_{\Lambda}$ & $-0.03 \pm 0.14$ & $-0.02 \pm 0.17$ \\ 
& $\Delta s_{\Lambda}$ & $0.74 \pm 0.17$ & $1.21 \pm 0.54$ \\ 
& $\Delta\Sigma_{\Lambda}$ & $0.68 \pm 0.44$ & $1.17 \pm 0.65$%
\end{tabular}
\end{table}

\end{document}